\pdfoutput=1
\documentclass[conference]{IEEEtran}
\IEEEoverridecommandlockouts
\usepackage{cite}
\usepackage{amsmath,amssymb,amsfonts}
\usepackage{graphicx}
\usepackage{textcomp}
\usepackage{booktabs}
\usepackage{multirow}
\usepackage{xcolor}
\usepackage{url}
\usepackage{subcaption}
\def\BibTeX{{\rm B\kern-.05em{\sc i\kern-.025em b}\kern-.08em
    T\kern-.1667em\lower.7ex\hbox{E}\kern-.125emX}}

\usepackage{stfloats}
\usepackage{float}
\usepackage{algorithm}
\usepackage{algpseudocode}
\usepackage{listings}
\usepackage{xcolor}

\definecolor{verylightgray}{RGB}{245,245,245}

\lstdefinestyle{promptstyle}{
    backgroundcolor=\color{verylightgray},
    basicstyle=\ttfamily\small,
    breaklines=true,
    columns=fullflexible,
    keepspaces=true,
    frame=single,
    rulecolor=\color{verylightgray},
    moredelim=**[is][\bfseries]{@@}{@@}
}

\begin{document}

\title{Code Comprehensor: Hierarchical Documentation \\ and Cross-Modal Retrieval for Large Repositories}

\author{\IEEEauthorblockN{Sujoy Roychowdhury\IEEEauthorrefmark{1},
        Joseph H. R. Isaac\IEEEauthorrefmark{2}, Jiangning Gao\IEEEauthorrefmark{3},
        Martin Klitte\IEEEauthorrefmark{4} and Diarmuid Corcoran\IEEEauthorrefmark{5}}
    \IEEEauthorblockA{Ericsson R\&D\\
        Email: \{\IEEEauthorrefmark{1}sujoy.roychowdhury,
        \IEEEauthorrefmark{2}joseph.isaac,
        \IEEEauthorrefmark{3}jiangning.gao,
        \IEEEauthorrefmark{4}martin.klitte,
        \IEEEauthorrefmark{5}diarmuid.corcoran\}@ericsson.com
    }}

\maketitle

\begin{abstract}
    Understanding large, complex, and rapidly evolving code bases remains a core challenge for software engineering at scale, hindering onboarding, maintenance, refactoring, and impact analysis. Providing this information to a Large-Language Model (LLM) is another hurdle, as the LLM must understand not only the required code method but also its associated methods in the grand scheme of the repository.
    We propose Code Comprehensor, a hybrid retrieval–reasoning framework combining (i) CodeGraph, a code dependency callgraph, (ii) CodeScribe, which uses code-aware LLMs to generate hierarchical documentation across repository call graphs, and (iii) CodeContrast, a CLIP-style dual-encoder model trained on CodeScribe output to learn aligned text-code embeddings. CodeContrast enables fast, semantically precise retrieval of relevant code-context neighborhoods, which serve as grounded evidence for an MCP-based LLM orchestrator answering domain-specific queries.
    Evaluation on OpenCV and a proprietary 5M-LOC repository shows that CodeScribe-derived supervision significantly improves cross-modal retrieval in text-to-code accuracy over baseline approaches. Richer textual supervision (hierarchical documentation, parent/sibling context, and LLM-generated synthetic questions) consistently enhances performance across both repositories.
    LLM-as-a-judge evaluation using the popular RAGAS of the Code Comprehensor output yields above 90\% faithfulness scores.
\end{abstract}


\section{Introduction}

Enterprises typically have large code repositories that have been developed and maintained over the years. The ability to document and understand such complex repositories has been a challenge \cite{cerny2020code, sayago2021systematic}. Code complexity creates dependencies on senior engineers' expertise to understand and navigate such repositories, and introduces inefficiencies in the Software Development Life Cycle (SDLC) \cite{hammad2020mining}. The growth of Large Language Models (LLM) in recent years and their ability to understand and explain code introduce the possibility of making this process more efficient \cite{macneil2023experiences}. This growth will lead to greater productivity in developing new features and maintaining and debugging existing ones.

However, the ability to use LLMs to document and understand large repositories in a reliable, replicable fashion requires careful design and consideration. Despite the large context window of many modern LLMs \cite{AwesomeAgentsLongContextLeaderboard2026}, code repositories often will consume a much larger number of tokens. For example, \cite{RepoStats2026} gives Lines of Code (LOC) for some public repositories, and token sizes will be much larger. Proprietary code repositories are often large in terms of LOC. For example, one of the repositories mentioned in this paper is larger than most listed in \cite{RepoStats2026}. More importantly, just because an LLM can process a large number of tokens does not mean an LLM should access the entire repository for every question. In addition to cost implications, large context sizes are known to give poorer results \cite{du2025context,firooz2024lost}.

Additionally, code repositories often contain identical or very similar files, methods, and variables across different parts of the code base, and using probabilistic machinery like an LLM is unlikely to be reliable or replicable. In view of this, it is important to design systems that enable LLMs to navigate code repositories more consistently and deterministically. Regarding this issue, some studies discuss using the inherent structure of code, as captured in the call graph, to drive documentation. Work by \cite{hoang2025codewiki} discusses using a static-analysis-based call tree to generate documentation for large-scale repositories recursively. A similar approach to generate initial and incremental documentation has been followed by other works \cite{luo2024repoagent}. A hierarchical engine for document generation and querying has also been studied \cite{sounthiraraj2025code}. Recently, Google released a repository to document code \cite{GoogleCodeWiki2025}; however, to the best of our knowledge, the system's design has not been disclosed. The common aspect of these repositories is the use of either the Abstract Syntax Tree (AST) \cite{zhang2019novel} or the call-graph. However, if one wants to build more than just static documentation, a broader and more comprehensive approach is needed. Previous work \cite{shi2022cross,li2023rethinking} has looked at cross-modal contrastive learning for code search; however, without deep domain-level documentation mapping methods to domain and code structure, the effectiveness of such cross-modal retrieval on large, complex repositories is limited.

\begin{figure*}[t!]
    \centering
    \includegraphics[width=0.8\textwidth]{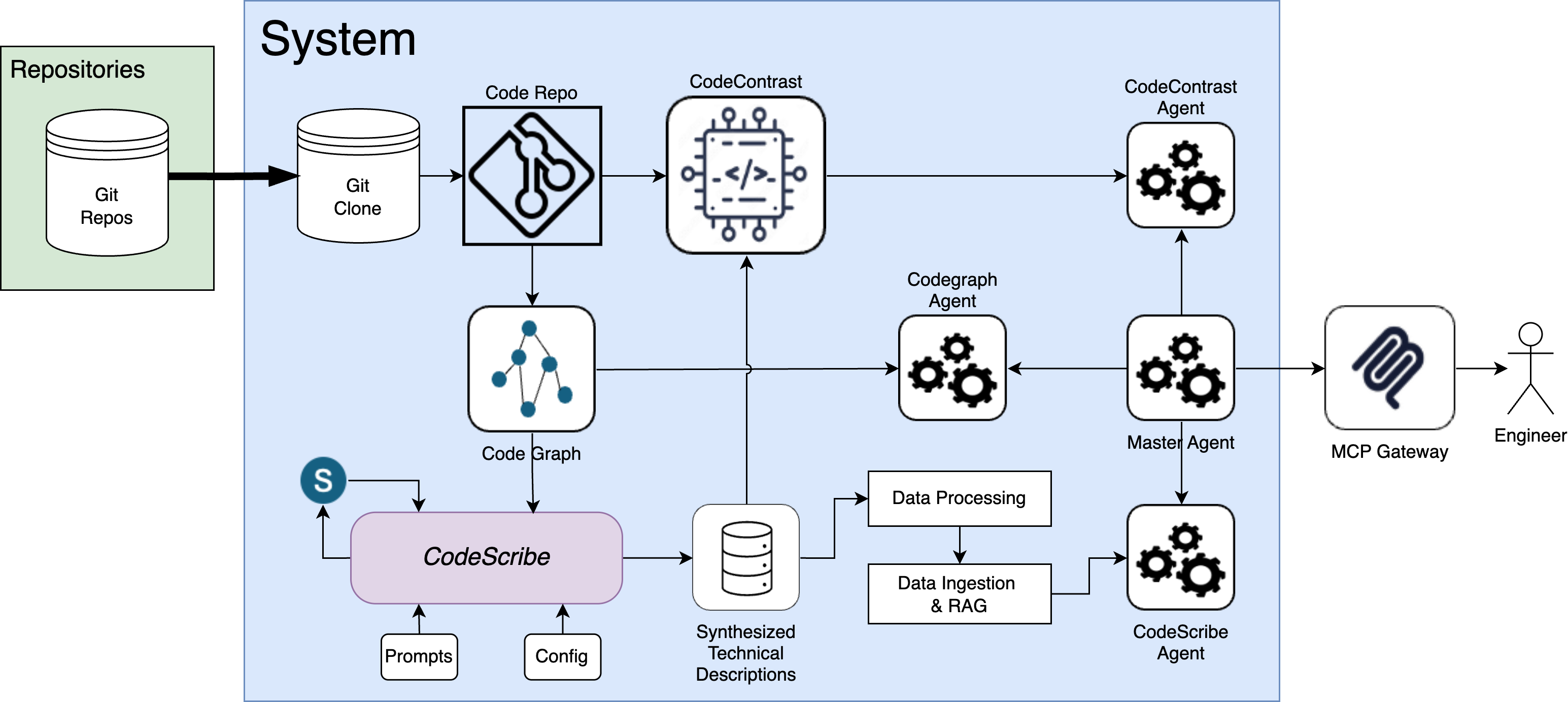}
    \caption{Overview of the Code Comprehensor architecture. CodeGraph is created through static analysis and consumed by CodeScribe to generate structured technical descriptions. These descriptions are used by the contrastive model for text and code retrieval. An agentic layer built on the Model Context Protocol coordinates CodeGraph traversal, CodeScribe based retrieval, and contrastive model queries for interactive use.}
    \label{fig:systemArch}
\end{figure*}

\subsection{Proposed Solution}

We present \textbf{Code Comprehensor}, an end-to-end system that combines four main components: (i) \textbf{CodeGraph}, a call graph generated via static analysis and exposed through API endpoints for traversal and chunking; (ii) \textbf{CodeScribe}, a hierarchical documentation engine that recursively processes directory-structured code, supports multiple documentation chapters and custom prompts per directory, and outputs PlantUML diagrams; (iii) a custom data processing module that parses CodeScribe output into relational records linking methods to parent/sibling context; and (iv) \textbf{CodeContrast}, a CLIP-based dual-encoder contrastive model using InfoNCE loss for asymmetric text-to-code and code-to-text retrieval with multi-positive learning support.

CodeScribe operates on a repository directory structure with configurable chapter-based perspectives (e.g., code description, algorithms, data flow). For each start method (in-degree zero in the method subgraph), CodeScribe retrieves the Depth-First-Search (DFS) traversal tree via CodeGraph, chunks it if necessary using post-order traversal to preserve hierarchy, and generates documentation. CodeScribe supports both full and incremental (git-commit-driven) documentation runs.

CodeContrast leverages the structured training data from CodeScribe parsing to train a retrieval model. We adopt CLIP's dual-encoder architecture for efficient pre-computed embeddings and asymmetric modalities. The InfoNCE objective learns well-separated embedding spaces and extends naturally to multi-positive scenarios.

Finally, a Model Context Protocol (MCP) based orchestration layer integrates CodeGraph, CodeScribe, and CodeContrast into an agentic system accessible via an Integrated Development Environment (IDE). This Master Agent invokes specialized agents (RAG over CodeScribe output, CodeGraph traversal, CodeContrast retrieval) to answer domain-specific queries on code repositories.

The \textbf{core contributions} of this paper are as follows:
\begin{itemize}
    \item A post-order chunking algorithm that enables LLM documentation of arbitrarily large call trees while preserving structural hierarchy.

    \item A multi-granularity training data construction pipeline that leverages parent-sibling relationships from documentation structure and LLM-generated synthetic questions for enhanced retrieval performance.

    \item Code Comprehensor: an integrated end-to-end system combining CodeGraph, CodeScribe, and CodeContrast components for large-scale repository understanding and retrieval.

    \item Empirical evidence that hierarchical documentation-derived supervision improves cross-modal code retrieval by at least 9\% Top-1 accuracy over baseline approaches.
\end{itemize}

The remainder of the paper is structured as follows. We briefly summarize previous work in Section \ref{sec:RelatedWork}. We discuss the System Architecture in Section \ref{sec:architecture}, the detailed methodology in Section \ref{subsec:Methodology}, and present results in Section \ref{sec:results}. We conclude with a discussion, threats to validity, and future works in Sections \ref{sec:discussion}, \ref{sec:threats}, and \ref{sec:futureWork}, respectively.

\section{Related Work}\label{sec:RelatedWork}

\subsection{Code Search and Retrieval}

Code search and retrieval has been an active area of research for understanding large repositories. Recent work has focused on neural approaches combining text and code embeddings. Sun et al.~\cite{Sun2022} presented context-aware code translation for code search, while Chai et al.~\cite{Chai2022} explored cross-domain deep code search using meta-learning techniques. Mathew and Stolee~\cite{Mathew2021} investigated cross-language code search, and Mao et al.~\cite{Mao2023} proposed self-supervised query reformulation for improved retrieval. More recently, Wang et al.~\cite{Wang2024} examined fusion of multiple code searchers, and Shi et al.~\cite{Shi2023} introduced CoCoSoDa for effective contrastive learning on code search tasks. Large-scale contrastive pre-training for code retrieval has been explored by Li et al.~\cite{Li2022}, who developed CodeRetriever. Comprehensive surveys on code search have been provided by Di Grazia and Pradel~\cite{DiGrazia2023} and Sun et al.~\cite{Sun2024}.

\subsection{Repository-Level Documentation Generation}

Documentation generation at repository scale has emerged as a critical need for understanding complex systems. Luo et al.~\cite{Luo2024} introduced RepoAgent, which combines static analysis with LLM-driven documentation. Hoang et al.~\cite{Hoang2025} developed CodeWiki, a framework for generating hierarchical documentation. Sounthiraraj et al.~\cite{Sounthiraraj2025} proposed a hierarchical engine for structured document generation and querying.

\subsection{Contrastive Learning for Code}

Contrastive learning has proven effective for learning aligned embeddings of code and text. Neelakantan et al.~\cite{Neelakantan2022} introduced text and code embeddings using contrastive methods. Li et al.~\cite{Li2022} leveraged contrastive learning in CodeRetriever for large-scale pre-training. Shi et al.~\cite{Shi2023} proposed CoCoSoDa with a momentum encoder design.

\subsection{Differentiation from Prior Work}

Code Comprehensor extends prior work in three key ways. First, unlike code search methods that operate on isolated code snippets (e.g., CodeRetriever, CoCoSoDa), our system first generates hierarchical documentation that captures inter-method relationships through call graph analysis and multi-perspective documentation generation, then trains a contrastive model on this enriched representation. Prior work on code retrieval has focused primarily on Python and Java; CodeRetriever and CoCoSoDa lack publicly available implementations for C/C++ repositories, limiting direct empirical comparison on the systems programming domain where our work is evaluated. Second, unlike RepoAgent, which focuses solely on documentation generation, our system uses the generated documentation as a structured training signal for a retrieval model, enabling efficient inference-time retrieval without requiring LLM calls per query. Third, we employ a post-order chunking algorithm to handle arbitrarily large call trees while preserving structural hierarchy, enabling scalable documentation and retrieval on repositories exceeding 5M lines of code.

\section{System Architecture}\label{sec:architecture}

Figure \ref{fig:systemArch} shows the overall system architecture for Code Comprehensor. Git repositories are first cloned into our system, and the CodeGraph is generated using static analysis. This CodeGraph is then consumed in CodeScribe, which uses two additional external information sources: prompts and configuration files. These are discussed in detail in Section \ref{sec:CodeScribe}. The output of CodeScribe, called ``Synthesized Technical Descriptions" (SynTD), is consumed by CodeContrast. CodeScribe uses recursive synthesis (shown as ``S" in Figure \ref{fig:systemArch}) indicated by the self-loop. To enable an end user (or an external system/agent) to interact with the documentation, multiple custom agents are built. The end user interacts with the Master Agent (MA) via an MCP gateway. The MA is an LLM-based orchestrator that calls multiple specialized agents. The CodeScribe Agent is an RAG-based system on the SynTD. There is a CodeContrast agent that can invoke API calls on the trained contrastive model. The CodeGraph agent can perform custom traversals on the CodeGraph via an exposed API.

Given a natural-language question from the user (an Engineer in this example), the MA retrieves relevant code blocks from the CodeContrast agent using Text$\rightarrow$Code-based retrieval. The MA retrieves the hierarchy of the relevant methods from the CodeGraph agent and obtains their documentation using the CodeScribe Agent. Finally, the MA composes a natural-language response to the end user and adds relevant diagrams and code blocks to provide a more comprehensive answer. The end user can interact further with the MA with follow-up questions. The Engineer can also ask questions, such as explanations of specific files, methods, or classes, and their relevance to the overall system. In such a case, MA uses the Code$\rightarrow$Text retrieval from the CM agent. It also identifies the method or file hierarchy from the CodeGraph agent and the documentation from the CodeScribe Agent. The MA decides which type of retrieval to invoke from the CodeContrast agent. This system allows for seamless end-to-end navigation and interpretation of the repository.

\section{Methodology} \label{subsec:Methodology}

\subsection{CodeGraph Creation}\label{subsec:callgraph}
To create the CodeGraph, a static analysis of the code repository is first performed. Static analysis identifies all code files and their corresponding methods. 
Dependencies are resolved via compile commands or, if unavailable, with reasonable heuristics. Node types are files and methods, and in object-oriented languages, other node types, such as classes, are added. Comments written by the developer for the methods are also extracted as attributes for the method nodes in CodeGraph, along with the method signatures. For methods without a developer-written comment, an LLM is prompted to generate one. The actual method body is also available as an attribute on the method. Since the same method can appear in multiple files, nodes are named ``\texttt{method\_name:::file\_name}" where ``:::" is chosen as the separator since it is unlikely to appear elsewhere. In object-oriented languages, nodes are named ``\texttt{method\_name:::file\_name}::::line\_number" to handle operator overloading. Method names include the namespace and class name, where relevant, as fully qualified names. The Tree-Sitter library \cite{TreeSitter2026} is used for parsing, though other approaches can be used. The CodeGraph generation, of course, needs (programming) language-specific parsing. The CodeGraph structure is generic, and the rest of the processing in the entire system is language-agnostic.

The CodeGraph is saved as a Python Networkx \cite{NetworkX2026} library object. This CodeGraph is also exposed through a FastAPI service \cite{FastAPI2026}. Several endpoints are exposed via the FastAPI service. Most are standard endpoints for graph traversal, e.g., Breadth-First-Search (BFS) and Depth-First-Search (DFS), along with the ``invokes" edge in both forward and reverse directions. The resultant CodeGraph and the endpoints are used to generate the documents using CodeScribe.

\subsection{CodeScribe}\label{sec:CodeScribe}

CodeScribe documentation is driven by the repository's directory structure because that is the natural way developers navigate it. The output of CodeScribe is Markdown files in a directory structure that replicates the code repository's directories. CodeScribe has configuration files that allow one to exclude irrelevant directories and include only file name extensions that are present. Every directory can have one or more perspectives to document the code, called ``chapters'' (e.g., code description, algorithms, data flow, interfaces). For each chapter, a custom prompt is generated, with flexibility to customize per directory. CodeScribe can also consume configuration information, such as abbreviation lists for specialized domains.

The system prompts are designed with three key principles: (1) LLM output is constrained to generate documents at hierarchical levels of abstraction, summarizing at appropriate semantic boundaries; (2) quoted code segments are marked with metadata (method name, file path, language) to enable downstream parsing and tracing back to source; (3) consistency across documentation perspectives ensures coherent multi-chapter views of the same code units. These design choices are critical for downstream parsing and contrastive model training.

When processing any directory, CodeScribe invokes CodeGraph to find all files in the directory and the methods they contain. Next, among the methods it identifies, start methods are those with an in-degree of zero in the method-only subgraph (a subgraph consisting only of methods). These methods are invoked by the external system (outside the scope of the current module graph). For each start method, the DFS traversal tree is retrieved via an endpoint on the CodeGraph API. The DFS traversal endpoint generates a key for each element in the DFS walk, and this information is included in the LLM's context. The context also includes the method definitions in the DFS tree. Thus, there is one Markdown output file per chapter and per start method. The documents for each start method are then combined for a summarized document at the directory level. Optionally, via a configuration file, summarizations for directory hierarchies are also done.

However, in large repositories, it is common to find trees that exceed the allowable context window, especially since method definitions are included. It is important to note that context sizes include output tokens, and there must be a reasonable allowance left for the LLM to generate sufficiently detailed documentation. This requires that the DFS walk trees be properly chunked. Hence, we build a chunking process that allows customizable chunk sizes. To handle such large trees, we implement an algorithm that uses recursive post-order traversal to process the DFS tree. For each node, we first process all children recursively, collecting any remaining chunks. We then check if the current node plus all child remainders fit within the primary context limit. If so, we merge them. If not, we finalize each child remainder as an independent chunk and keep only the current node. Finally, if the resulting chunk exceeds the primary limit plus a buffer, we split it into overflow chunks and return the remaining tail.

The chunks are presented to the LLM as context, and the output of the previous chunk is added to future chunks in the context in a recursive generation process. The output of intermediate chunks is temporary and not included in the final documentation. Our algorithm ensures that when the LLM processes a parent caller, it possesses a high-level functional summary of all callee implementations, even if the source code of those children is not included in the current active window. This rolling summarization effectively compresses the global DFS walk state, allowing the model to generate accurate parent-level documentation without having to guess the behavior of its dependencies.

\begin{figure}[t!]
    \centering
    \includegraphics[width=0.5\textwidth]{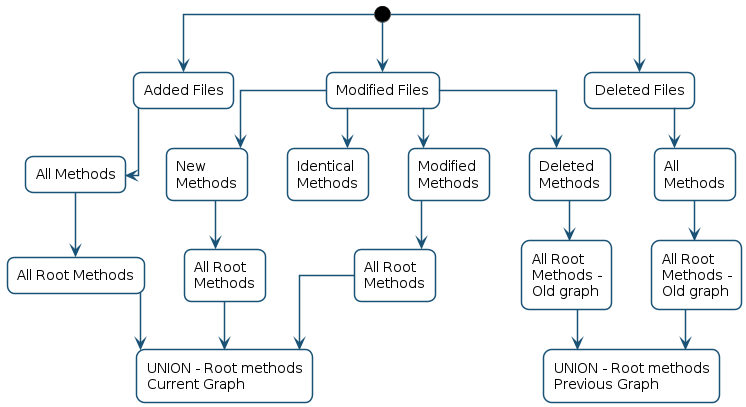}
    \caption{Incremental processing to identify all impacted root methods}
    \label{fig:incremental}
\end{figure}

CodeScribe can also process incrementals based on git commit IDs. This process can be done by identifying the last processed git commit and identifying files that have changed. Further to this, the methods in these impacted files are identified and checked against the previous version of the graph to see if they have changed. If they have, then the root/start methods of the trees they belong to are identified. This method is shown in Figure \ref{fig:incremental}. Once this is done, CodeScribe asks the LLM to modify only the start methods and any summarization documents generated for this. This process allows incremental processing at any frequency, from periodic intervals such as weekly to integration with Continuous Integration/Continuous Development (CI/CD) pipelines. CodeScribe's output is then parsed into a more palatable data format for CodeContrast.

\subsection{Parsing CodeScribe Data}\label{sec:parsing}
As mentioned before, CodeScribe generates Markdown output that mimics the repository's directory structure. The documents contain a mixture of the following:

\begin{itemize}
    \item Hierarchical section headers denoting algorithms and module structure.
    \item Inline code blocks annotated with method names, file paths, and programming language identifiers.
    \item PlantUML diagram specifications embedded within fenced code blocks.
    \item Free-form textual descriptions providing domain context, algorithmic rationale, and inter-component relationships.
\end{itemize}
To utilize this data for training CodeContrast, the Markdown files undergo several processes:

\subsubsection{Markdown-to-JSON Conversion}
A custom recursive parser was implemented to convert Markdown documents into nested JSON structures. The headers and subheaders of the Markdown file become the keys of the JSON file. Non-header content is aggregated under an arbitrary key at the appropriate nesting level. This approach preserves the hierarchical organization of the original documentation while enabling programmatic traversal.

\subsubsection{Code Block and Text Separation}
Each leaf node in the resulting JSON tree is processed to separate the code blocks from the text segments. A regex-based extractor identifies fenced code blocks (delimited by triple backticks) and captures both the language identifier and the source code. The remaining text, after code block removal, is retained as a description. The extraction procedure traverses the JSON tree recursively, assigning a dot-separated depth identifier (e.g., 1.2.3) to each node. This depth encoding enables the subsequent identification of parent-child and sibling relationships among extracted entries, which is critical for constructing the relational features of the final dataset.

\subsubsection{Code Block Classification}
Each extracted code block is classified into one of three categories based on its content:

\begin{itemize}
    \item \textbf{\textit{Code}}: Blocks containing annotated source code with explicit method name and file path metadata (extracted via inline comments from CodeScribe).
    \item \textbf{\textit{UML}}: Blocks identified as PlantUML specifications, detected by the presence of @startuml directives or the PlantUML language identifier.
    \item \textbf{\textit{Pseudocode}}: Code blocks lacking method name and file path annotations, treated as pseudocode or algorithmic sketches.
\end{itemize}

\subsubsection{Relational Feature Construction}
For each code entry, the pipeline constructs a rich set of relational features by using the depth-encoded tree structure from CodeScribe:

\begin{figure*}[]
    \centering
    \begin{minipage}{0.9\textwidth}
    \begin{lstlisting}[style=promptstyle]
@@Reference context@@(do NOT copy reference code verbatim into the questions):
Function signature (function name removed): bfloat16x8_t (float * src)
@@Comment@@: Loads 8 float values from memory and converts them to bfloat16 ...
@@Code snippet@@ (terminology only; do NOT copy):
bfloat16x8_t vld1q_as_bf16(const float* src){
    float32x4_t s0 = vld1q_f32(src), s1 = vld1q_f32(src + 4);
    return vcombine_bf16(vcvt_bf16_f32(s0), vcvt_bf16_f32(s1));}

@@BEHAVIOR_ANCHORS@@ (choose >=1 per question): convert, pack, register
@@PARAM_TERMS@@ (use in questions): src, float
@@CONTEXT_TERMS@@ (use in questions): NEON, ARM, SIMD, memory, bfloat16, vector, ...
@@KEY_TERMS@@ (use in questions): src, float, NEON, ARM, SIMD, memory, bfloat16, vector, ...

@@Task:@@
Generate EXACTLY 3 DISTINCT natural-language questions.
Hard constraints for each question:
- MUST include at least ONE BEHAVIOR_ANCHOR (verbatim).
- MUST include at least ONE term from PARAM_TERMS and at least ONE term from CONTEXT_TERMS.
- MUST include at least 3 items from KEY_TERMS (verbatim).
- MUST end with a question mark '?'.
- MUST NOT mention any function name, file path, any id, or file:line identifiers.

@@OUTPUT QUESTIONS@@: [
    "How can I convert eight consecutive float values from memory into a bfloat16 NEON vector using ARM SIMD instructions that maintains high precision bits?",
    "What ARM NEON approach should I use to pack multiple float values from src memory into a single bfloat16 vector through vectorized type conversion?",
    "How do I register-load float data from memory and convert them to bfloat16 format using NEON SIMD instructions for efficient floating point processing?"]
\end{lstlisting}
    \end{minipage}
    \caption{Example prompt for synthetic question generation and corresponding generated questions in OpenCV. Some words/phrases are in bold here for
        clarity. }
    \label{lst:synthq_prompt}
\end{figure*}

\begin{itemize}
    \item \textbf{\textit{Parent context}}: Textual descriptions from the immediate parent node, providing high-level context.
    \item \textbf{\textit{Sibling context}}: Method names, comments, file paths, and code of entries sharing the same parent, capturing co-located functionality.
    \item \textbf{\textit{Module-level context}}: Aggregated commentary from text-only nodes that lack direct child code blocks, serving as overarching module documentation.
\end{itemize}

These relational features are designed to provide the contrastive learning model with multi-granularity contextual signals linking code to its surrounding documentation. The parsed records are merged with the call graph data created earlier, using the same composite key of the form ``\texttt{method\_name:::file\_path}"

\subsection{Synthetic Data Generation}
\label{subsec:synthetic_data_generation}
Large repositories often have sparse, inconsistent, or outdated data (e.g., missing comments, legacy documentation, or domain-specific abbreviations). Moreover, real developer queries often take the form of questions rather than direct text descriptions. Therefore, to improve coverage and better match downstream retrieval and reasoning workloads, synthetic questions generated from the parsed records are augmented, with code included only as a grounding aid.

\subsubsection{Design Principles}
Synthetic questions are constructed with two explicit objectives:
\begin{itemize}
    \item \textbf{\textit{Intent diversity}}: cover common developer intents such as purpose/behavior, parameter usage, edge cases, and preconditions, error handling, and interactions within a local module context.
    \item \textbf{\textit{Retrieval grounding}}: include repository-specific terminology (parameter names, types, macros, struct/enum identifiers, and field names) to improve retrievability, while explicitly discouraging ``keyword dumping''.
\end{itemize}
In our pipeline, the \emph{semantic intent} of each question is primarily driven by the available documentation signal. At the same time, the method signature and a bounded code snippet are used as auxiliary sources for terminology anchors.

\subsubsection{Generation Inputs}
Each synthetic question set is generated for a single \textbf{\textit{code entry}} produced by Section~\ref{sec:parsing}.

For a given entry, the generation payload consists of:
\begin{itemize}
    \item \textbf{\textit{Documentation signal}}: one consolidated text slice corresponding to a selected comment level or hierarchical comment fields when available.
    \item \textbf{\textit{Code signal}}: the method signature and a bounded code snippet (provided for terminology only, with explicit instructions not to copy verbatim). We provide a signature with the symbol name removed.
\end{itemize}
The generator, therefore, receives a comment-driven description of behavior, augmented with a limited amount of code-local context.

\subsubsection{Terminology Anchor Extraction}
To encourage grounded and retrievable questions, we extract candidate anchors from the entry:
\begin{itemize}
    \item \textbf{\textit{Behavior anchors}} (\texttt{BEHAVIOR\_ANCHORS}): action-level intent tokens (verbs) that constrain question semantics.
    \item \textbf{\textit{Parameter anchors}} (\texttt{PARAM\_TERMS}): parameter names and salient type tokens parsed from the function signature.
    \item \textbf{\textit{Context anchors}} (\texttt{CONTEXT\_TERMS}): repository- and domain-specific terminology mined from the documentation slice and the bounded code context.
    \item \textbf{\textit{Key terms}} (\texttt{KEY\_TERMS}): a capped union of parameter and context anchors, used as a lexical grounding constraint while limiting the total number of enforced tokens to avoid keyword stuffing.
\end{itemize}

\subsubsection{Prompting Strategy: Comment-Driven Intent with Anchored Questions}
We use a code-capable instruction-following LLM to generate synthetic questions.\footnote{We use \texttt{[GEN\_MODEL]} with decoding parameters \texttt{[TEMP]}, \texttt{[TOP\_P]}, and \texttt{[MAX\_TOKENS]}.}

For each entry, the prompt instructs the generator to produce developer-style questions with the following rules:
\begin{itemize}
    \item Reflect the described behavior and constraints from the documentation signal;
    \item includes at least two terminology anchors (preferably one from \texttt{PARAM\_TERMS} and one from \texttt{CONTEXT\_TERMS} when available), and at least three verbatim hits from \texttt{KEY\_TERMS};
    \item Include at least one \texttt{BEHAVIOR\_ANCHORS} (e.g., \emph{validate}, \emph{initialize}, \emph{convert}, \emph{decode}) to reduce purely token-driven outputs, while discouraging keyword dumping;
    \item Enforce interrogative form (each question ends with `?');
    \item Avoid copying code verbatim and avoid dataset-specific identifiers (e.g., \texttt{<path>:<line>} strings).
\end{itemize}

This design yields questions whose intent is comment-driven, while code-derived anchors improve grounding and retrieval utility. Listing~\ref{lst:synthq_prompt} shows an example prompt used for synthetic question generation. We instruct the model to produce exactly three developer-style questions for each method.

\begin{figure*}[!t]
    \centering
    \includegraphics[width=0.75\linewidth]{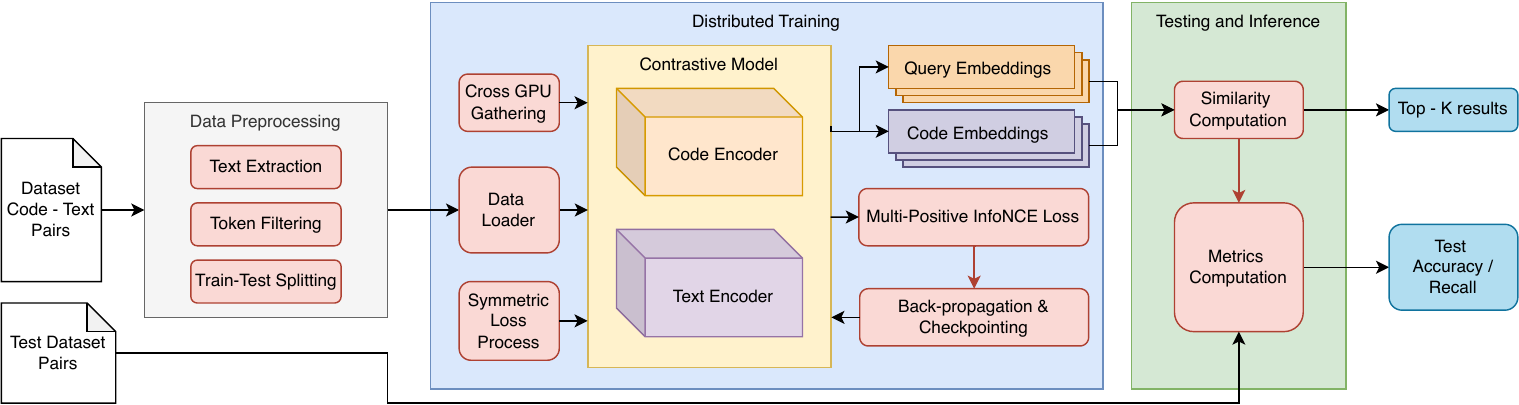}
    \caption{Overview of the CodeContrast workflow, covering data preprocessing, distributed training, along with the testing and inference processes.}
    \label{fig:architecture}
\end{figure*}

\subsubsection{Post-processing}
We apply lightweight filtering to improve precision and reduce noise:
\begin{itemize}
    \item \textbf{\textit{Format and alignment checks}}: enforce strict JSON output and validate echoed tags; discard malformed outputs.
    \item \textbf{\textit{Deduplication}}: remove exact and near-duplicate questions within the same ID.
    \item \textbf{\textit{Informativeness}}: discard questions that do not contain the required anchors or fall below a minimum length.
\end{itemize}
To assess retrieval generalization beyond trivial lexical shortcuts, we additionally evaluate a post-processed variant where function-name tokens are removed from questions.

\subsection{CodeContrast Architecture}

CodeContrast is a custom dual-encoder contrastive learning system \cite{bai2025dual} that aligns natural-language documentation embeddings with source-code embeddings. The system is designed for a code retrieval task: given a textual query (e.g., documentation or comments), retrieve the most relevant code snippet from a corpus, and vice versa. It uses PyTorch's Fully Sharded Data Parallel (FSDP) \cite{ZhaoFSDP} for distributed multi-GPU training, supports both SentenceTransformer (SBERT) \cite{reimers2019sentence} and normal BERT Transformer \cite{devlin2019bert} backbones, and includes a full pipeline from data preparation through training, embedding store creation, and evaluation. The full architecture is shown in Figure~\ref{fig:architecture}.

\subsubsection{Data Preparation Pipeline}
This module serves as the initial stage of the data‑processing pipeline. It loads a JSON input file that is created from Section~\ref{sec:parsing}, and applies validation and filtering logic to produce structured entries containing code, documentation, unique identifiers, and query types. This induces a \emph{one-to-many} supervision structure: one code unit corresponds to many valid textual descriptions:
\begin{itemize}
    \item a single \textbf{\textit{code}} unit (method signature and bounded code snippet), and
    \item multiple \textbf{\textit{text views}} (raw documentation, hierarchical context, and synthetic questions), each tagged by a \texttt{query\_type}.
\end{itemize}

During this step, the system also flags entries that may cause issues, such as those with insufficient documentation or code that is too long to process effectively. Once everything is cleaned and organized, the module splits the data into training, validation, and test sets and then saves the final outputs.

\subsubsection{Core Model Architecture}
The embedding system supports several backbone options. In one setup, we use SBERT encoders, which produce sentence embeddings directly. In another, we use Transformer models such as BERT, where we create fixed-size representations either by averaging the contextual token embeddings or by taking the final token embedding. To make sure these different backbones produce vectors of the same size, we optionally add a projection layer that maps everything to a shared dimension $d$. All embeddings are then $\ell_2$-normalized onto the unit hypersphere, enabling cosine similarity to be computed by dot product.

A dual‑encoder framework integrates these embedding models in a unified contrastive learning setup. Each encoder processes either a text query or a code snippet, and both are trained jointly using a similarity objective that is adjusted by a temperature parameter. Because training runs across several devices, we gather embeddings from all GPUs to form a single global similarity matrix. This lets the model compare every item in the batch to all others, increasing the number of negative examples and effectively creating a much larger batch. In practice, this leads to more stable training and stronger contrastive representations.

\subsubsection{Differentiable Global-Batch Training under FSDP}
We use the Fully Sharded Data Parallel (FSDP) paradigm for training - this runs across multiple devices, so we bring all embeddings and labels together into one global batch before computing the similarity matrix. This gives the model a much larger set of negatives and keeps the contrastive comparisons consistent across GPUs. To make this work, we use a differentiable all‑gather operation that allows gradients to flow through the gathered tensors, which keeps the entire contrastive objective fully trainable even in a distributed setup. For safety, any rank that produces an empty batch still participates in the gather using placeholder tensors. This avoids synchronization issues and prevents the training process from stalling. 

\subsubsection{Multi-positive Symmetric Contrastive Objective}
Multiple text views may describe the same underlying code unit, and these code-text pairs may appear multiple times within a batch, depending on sampling. Therefore, positives using identifier labels are defined. Let $\ell_i$ denote the class label derived from $\texttt{id}_i$. For each text anchor $t_i$, the set of positive code indices is
$P(i)=\{j \mid \ell_j=\ell_i\}$.
The Text$\rightarrow$Code loss uses a multi-positive InfoNCE objective with log-sum-exp aggregation:
\begin{equation}
    \label{eq:mp_t2c}
    \mathcal{L}^{t\rightarrow c}
    =
    -\frac{1}{B}\sum_{i=1}^{B}
    \log
    \left(
    \frac{\sum_{j\in P(i)} \exp(\mathbf{S}_{ij})}
    {\sum_{j=1}^{B} \exp(\mathbf{S}_{ij})}
    \right).
\end{equation}
Similarly, for Code$\rightarrow$Text we swap roles and apply the same objective to $\mathbf{S}^\top$:
\begin{equation}
    \label{eq:mp_c2t}
    \mathcal{L}^{c\rightarrow t}
    =
    -\frac{1}{B}\sum_{j=1}^{B}
    \log
    \left(
    \frac{\sum_{i: \ell_i=\ell_j} \exp(\mathbf{S}_{ij})}
    {\sum_{i=1}^{B} \exp(\mathbf{S}_{ij})}
    \right).
\end{equation}
We optimize a symmetric objective:
\begin{equation}
    \label{eq:sym}
    \mathcal{L}=\frac{1}{2}\left(\mathcal{L}^{t\rightarrow c}+\mathcal{L}^{c\rightarrow t}\right).
\end{equation}

\subsubsection{Evaluation Metrics}
We evaluate retrieval performance in both directions using metrics derived from the cross-modal similarity matrix $\mathbf{S}$ . For each query, we define the top matches as the candidates with the highest matching similarity scores, and report Top-$k$ accuracy. Recall@K measures how many of the relevant items are successfully retrieved within the topK results out of the total number of relevant items.
We also report direction-specific metrics by evaluating both $\mathbf{S}$ (Text$\rightarrow$Code) and $\mathbf{S}^\top$ (Code$\rightarrow$Text).

\section{Results} \label{sec:results}

\subsection{Test Methodology}
The objective of the testing phase is to assess how effectively the model retrieves the correct code function when provided with a natural language query (Text$\rightarrow$Code), and how well it identifies the appropriate textual description when given a code snippet (Code$\rightarrow$Text). The evaluation uses a held‑out test set consisting of methods paired with their corresponding text–code samples. Each sample includes a function along with all associated textual descriptions, such as raw docstrings, brief comments, self‑comments, and synthetic questions. None of the test methods or their synthetic variants are included in the training data, which ensures fair and unbiased assessment. Two types of tests are conducted. The first is the non‑synthetic test, which pairs each code sample with its original human‑authored (or LLM-authored when absent) textual descriptions. The second is the synthetic test, which augments each original pair with LLM-generated questions derived from the same function, as described in Section~\ref{subsec:synthetic_data_generation}.

In addition to testing for the accuracy of CodeContrast, we also do an evaluation for testing the output of the Code Comprehensor system by using LLM-as-a-judge for the output of the MA. As this is a subjective evaluation, we use the RAGAS framework \cite{es2024ragas}. In particular, we use the Answer Correctness (AC) and Faithfulness (FF) metrics of this framework; earlier studies have shown that these have the best correlation with ground truth \cite{roychowdhury10evaluation}. This evaluation is done on a subset of the synthetic questions; the generated answer from the MA is compared with the ground-truth answer used at the time the synthetic questions were generated.

\subsection{Dataset}
We evaluate our results on two repositories. One of them (called Repo-A) is a proprietary C-language repository in a large telecommunications organization, having over 5 million Lines of Code (LOC) in ``.c" and ``.h" files only. In addition, to ensure replicability, we run our experiments using OpenCV\footnote{\url{https://github.com/opencv/opencv}}. OpenCV has about 1.2 million LOC across its ".c", ".h", and ".cpp" files. We use Anthropic Sonnet 4.5 as the LLM for CodeScribe and Anthropic Sonnet 4.0 for all other generation tasks (missing comments, synthetic questions).

\subsection{CodeScribe output on OpenCV}

CodeScribe output quality can be verified subjectively. Subject Matter Experts (SMEs) have verified the results of the proprietary repository and have expressed satisfaction with the output. For OpenCV, the authors did a manual comparison themselves, as they are quite conversant with the library.

\begin{figure*}[]
    \centering
    \includegraphics[width=0.95\textwidth]{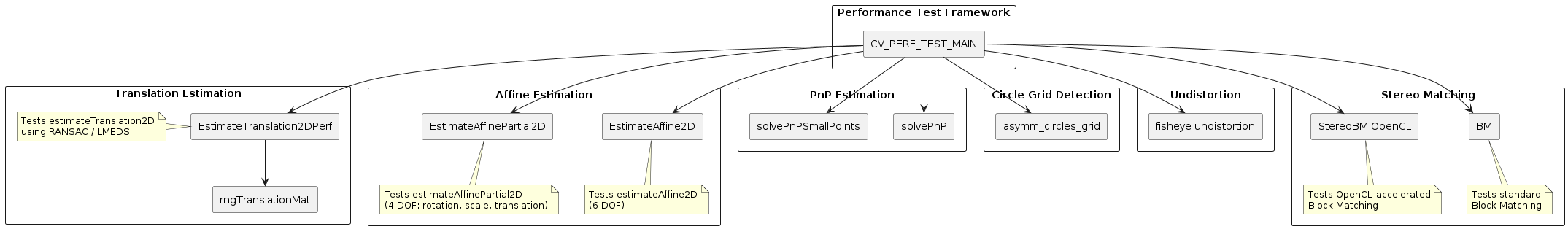}
    \caption{Sample diagrammatic representation of sub-modules in OpenCV Calib3d module. This module provides 3D calibration features and has a number of important submodules.}
    \label{fig:opencv}
\end{figure*}

We compare the output with CodeWiki \cite{GoogleCodeWiki2025}, which is publicly available for OpenCV.
For example, in the section on 3D calibration, one of the important features available in OpenCV, CodeScribe provides a modular representation of its important components and discussions on each of them separately. This is shown in Figure \ref{fig:opencv}, which is much more informative than the corresponding module\footnote{\url{https://codewiki.google/github.com/opencv/opencv#core-modules-and-algorithms-camera-calibration-and-3d-reconstruction}}  in CodeWiki.

The documentation is largely correct and useful, but somewhat verbose. This documentation may be very helpful for an agent to consume; in some cases, it can be challenging for humans to browse. However, the Master Agent can interactively expand or summarize the documentation, along with additional context from the other agents. Additionally, CodeScribe generates component diagrams that provide a systematic understanding of the repository's contents.

\subsection{Computational Efficiency and Costs}

The practical deployment of Code Comprehensor requires an understanding of resource requirements. CodeScribe generation for OpenCV (1.2M LOC) required approximately 48 hours and \$3,200 USD in LLM API costs using Sonnet 4.5. For Repo-A (5M LOC), the generation required approximately 8 days and a few thousand USD. CodeContrast training on 4 NVIDIA A100 GPUs converged in 15 epochs over approximately 24 hours. Inference is significantly more efficient: CodeContrast retrieval requires only a forward pass through the text encoder followed by cosine similarity computation against pre-computed code embeddings, achieving $<$50 ms latency per query. The full Master Agent pipeline (text-to-code retrieval + context assembly + LLM response generation) averages 3-5 seconds per query, depending on the result set size and the complexity of the response synthesis. These costs are one-time for initial documentation generation and model training; incremental updates (triggered by code changes) cost significantly less per month.

\begin{table*}[t]
    \centering
    \caption{Top-$K$ retrieval results across four model configurations.
        Top-$K$ columns report text$\rightarrow$code accuracy and code$\rightarrow$text accuracy;
        Recall@K (c$\rightarrow$t) reports recall for the code$\rightarrow$text task. $N$ denotes the number of test samples for results in those rows.}
    \label{tab:topk_opencv_configs}
    \resizebox{\textwidth}{!}{
        \begin{tabular}{cccccccccccccccc}
            \toprule
            \multirow{2}{*}{Repository}                           &
            \multirow{2}{*}{Configuration}                        &
            \multirow{2}{*}{Test Type}                            &
            \multicolumn{4}{c}{TopK (text$\rightarrow$code) $\%$} &
            \multicolumn{4}{c}{TopK (code$\rightarrow$text) $\%$} &
            \multicolumn{4}{c}{Recall@K (c$\rightarrow$t) $\%$}                                                                                                                                                                                                                                                                                      \\
            \cmidrule(lr){4-7} \cmidrule(lr){8-11} \cmidrule(lr){12-15}
                                                                  &                                &                                     & Top1           & Top3           & Top5           & Top10
                                                                  & Top1c                          & Top3c                               & Top5c          & Top10c
                                                                  & R@1                            & R@3                                 & R@5            & R@10                                                                                                                                                                                     \\
            \midrule
            \multirow{4}{*}{OpenCV}                               & D1                             & \multirow{1}{*}{Non-synthetic test} & 92.68          & 97.23          & 97.94          & 98.36          & 91.99          & 97.62          & 98.04          & 98.51          & 91.04          & 97.41          & 97.94          & 98.50          \\
                                                                  & D2                             & N = 2813                            & \textbf{94.45} & \textbf{98.19} & \textbf{98.54} & \textbf{99.15} & \textbf{94.66} & \textbf{97.97} & \textbf{98.50} & \textbf{99.00} & \textbf{93.42} & \textbf{97.75} & \textbf{98.41} & \textbf{98.98} \\
            \cmidrule(lr){2-15}
                                                                  & D2                             & \multirow{1}{*}{Synthetic}          & 66.05          & 80.94          & 85.23          & 89.37          & \textbf{96.53} & \textbf{99.28} & \textbf{99.50} & \textbf{99.61} & \textbf{95.17} & \textbf{99.02} & \textbf{99.43} & \textbf{99.59} \\
                                                                  & D3                             & N = 12658                           & \textbf{80.64} & \textbf{93.29} & \textbf{95.83} & \textbf{97.61} & 93.67          & 98.07          & 98.68          & 98.82          & 92.24          & 97.82          & 98.57          & 98.80          \\
            \midrule
            \midrule

            \multirow{4}{*}{Repo-A}                               &
            D1
                                                                  & \multirow{1}{*}{Non-synthetic} & 74.82                               & 89.92          & 93.31          & 95.15
                                                                  & 74.21                          & 89.12                               & 92.47          & 95.21
                                                                  & 67.91                          & 86.34                               & 91.40          & 95.07                                                                                                                                                                                    \\

                                                                  & D2
                                                                  & N = 3800                       & \textbf{79.54}                      & \textbf{92.72} & \textbf{94.82} & \textbf{96.23}
                                                                  & \textbf{79.28}                 & \textbf{92.65}                      & \textbf{94.85} & \textbf{96.36}
                                                                  & \textbf{72.42}                 & \textbf{90.02}                      & \textbf{93.95} & \textbf{96.29}                                                                                                                                                                           \\
            \cmidrule(lr){2-15}

                                                                  & D2
                                                                  & \multirow{1}{*}{Synthetic}     & 67.46                               & 84.21          & 88.16          & 91.69
                                                                  & 80.41                          & 92.44                               & 94.75          & 96.44
                                                                  & 74.37                          & 90.01                               & 93.93          & 96.32                                                                                                                                                                                    \\

                                                                  & D3
                                                                  & N = 23480                      & \textbf{76.18}                      & \textbf{91.20} & \textbf{93.91} & \textbf{96.01}
                                                                  & \textbf{81.24}                 & \textbf{93.56}                      & \textbf{96.05} & \textbf{97.43}
                                                                  & \textbf{75.43}                 & \textbf{91.10}                      & \textbf{95.13} & \textbf{97.34}                                                                                                                                                                           \\
            \bottomrule
        \end{tabular}}
\end{table*}

\begin{table*}[t]
    \centering

    \textbf{Query-Type Definitions:}
    \begin{itemize}
        \item \texttt{self\_comment}: The CodeScribe-generated description of the method itself (self-level documentation).
        \item \texttt{doc\_brief\_merged}: The original developer-written docstring/brief comment merged with any existing documentation.
        \item \texttt{qa\_doc\_gen}: Synthetic questions generated from CodeScribe-produced documentation.
        \item \texttt{qa\_doc\_raw}: Synthetic questions generated from the raw/original developer documentation.
        \item \texttt{qa\_self}: Synthetic questions generated from the self-comment level documentation.
        \item \texttt{qa\_parent}: Synthetic questions generated from parent-level context documentation.
        \item \texttt{qa\_sibling}: Synthetic questions generated from sibling-level context documentation.
    \end{itemize}

    \vspace{6pt}

    \caption{Per–query-type Top-$K$ (text$\rightarrow$code) accuracy comparing \textbf{D2} vs \textbf{D3} for the OpenCV Repository.}
    \label{tab:topk_bytype_text2code}
    \small
    \renewcommand{\arraystretch}{1.1}
    \begin{tabular}{lccccccccc}
        \toprule
        \multirow{2}{*}{Query Type}               &
        \multirow{2}{*}{Number of Samples (Test)} &
        \multicolumn{4}{c}{D2 (TopK) $\%$}        &
        \multicolumn{4}{c}{D3 (TopK) $\%$}                                                                                                                                                       \\
        \cmidrule(lr){3-6}\cmidrule(lr){7-10}
                                                  &      & Top1           & Top3           & Top5           & Top10          & Top1           & Top3           & Top5           & Top10          \\
        \midrule
        \texttt{self\_comment}                    & 209  & 65.55          & 80.38          & 84.69          & 88.04          & 65.55          & \textbf{84.69} & \textbf{88.99} & \textbf{93.30} \\
        \texttt{doc\_brief\_merged}               & 2804 & \textbf{96.43} & \textbf{99.14} & \textbf{99.36} & \textbf{99.54} & 93.94          & 98.18          & 98.64          & 98.82          \\
        \texttt{qa\_doc\_gen}                     & 8049 & 57.39          & 76.43          & 82.02          & 87.25          & \textbf{78.38} & \textbf{93.14} & \textbf{96.15} & \textbf{98.05} \\
        \texttt{qa\_doc\_raw}                     & 351  & 78.06          & 94.02          & 94.87          & 97.72          & \textbf{83.19} & \textbf{97.72} & \textbf{99.43} & \textbf{99.43} \\
        \texttt{qa\_self}                         & 418  & 62.20          & 78.47          & 83.25          & 90.67          & \textbf{76.79} & \textbf{94.02} & \textbf{96.89} & \textbf{98.56} \\
        \texttt{qa\_parent}                       & 144  & 54.17          & 70.14          & 75.69          & 79.17          & \textbf{69.44} & \textbf{88.19} & \textbf{91.67} & \textbf{93.75} \\
        \texttt{qa\_sibling}                      & 402  & 44.78          & 60.45          & 67.16          & 74.38          & \textbf{63.93} & \textbf{78.86} & \textbf{84.83} & \textbf{90.80} \\
        \bottomrule
    \end{tabular}
\end{table*}

\subsection{Configurations}
To study how different levels of natural language supervision influence cross-modal retrieval, we define three dataset configurations, D1 through D3. Each configuration adds an additional layer of textual descriptions to the same code elements. This gradual buildup allows us to clearly observe how various comment types and question-answer styles affect the model's ability to generalize.

\subsubsection{D1 — Method + Docstring (Baseline Model)}
This configuration uses only the method's source code paired with its corresponding docstring or brief functional description. Docstrings provide a concise, developer‑written explanation of what the method does and form the foundational text modality for contrastive learning. They represent the simplest and most common documentation layer in code repositories.

\subsubsection{D2 — Method + Docstring + Self‑Comments}
D2 augments D1 by adding self‑comments, which were extracted from the CodeScribe data (Refer Section \ref{sec:parsing}). These comments offer finer‑grained insights into the how and why of the code's behavior. The model is then trained on text that will have self‑comments added to the text corpus.

\subsubsection{D3 — D2 + Parent–Sibling Comments and QA (Structured Context Expansion)}
D3 introduces higher‑level relational documentation by adding parent and sibling comments extracted from the CodeScribe output. This stage also includes QA derived from these comment types, enabling the model to learn abstractions that span across method boundaries and reflect multi‑method interactions. D3, therefore, represents the richest supervision setting, capturing both local method semantics and global contextual semantics.

\begin{table}[h]
\centering
\caption{Mean scores (\%) for evaluation of master agent using RAGAS metrics. AC: Answer Correctness}
\label{tab:agenticEvals}

\resizebox{0.65\columnwidth}{!}{%
\begin{tabular}{lcc}
\hline
Metric & Repo-A & OpenCV \\
\hline
No. samples & 4986 & 4971 \\
AC: Precision & 72.0 & 87.5 \\
AC: Recall & 58.3 & 72.1 \\
AC: F1 & 60.0 & 76.4 \\
Faithfulness & 94.6 & 93.4 \\
\hline
\end{tabular}%
}
\end{table}

\subsection{Evaluation of Master agent}

Table \ref{tab:agenticEvals} presents the agentic evaluation results for the MA using RAGAS metrics. The high FF scores indicate strong grounding of the generated responses in the retrieved context. The comparatively lower AC scores are influenced by the limited number of claims in both the ground-truth and generated responses, which results in a small denominator during metric computation and increases score sensitivity to individual mismatches.

\section{Discussion} \label{sec:discussion}
The results across both repositories demonstrate the nuanced benefits of incorporating CodeScribe-generated information into the CodeContrast pipeline. In Table~\ref{tab:topk_opencv_configs}, the Top-1 accuracy overall for OpenCV is 66.05\% for D2 and is 80.64\% for D3, showing an improvement of 14.59\%. Similarly, for Repo-A, we get 8.72\% improvement (67.6\% for D2 and 76.18\% for D3). In Table~\ref{tab:topk_bytype_text2code}, for non-synthetic tests on human-authored descriptions, D2 achieves superior performance: on OpenCV's \texttt{doc\_brief\_merged} category, D2 achieves Top-1 accuracy of 96.43\% while D3 achieves 93.94\%. This indicates that adding parent/sibling context and synthetic QA can reduce performance when evaluating against developer-authored comments. 

However, for synthetic test queries, D3 demonstrates consistent improvements over D2 across all Top-$K$ values for complex query categories (\texttt{qa\_doc\_gen}, \texttt{qa\_doc\_raw}, \texttt{qa\_self}, \texttt{qa\_parent}, \texttt{qa\_sibling}). This improvement compensates for the drop in accuracy in \texttt{doc\_brief\_merged} discussed earlier.
To ensure that function names in synthetic questions do not lead to data leakage, we remove function-name tokens from questions - the results in Table~\ref{tab:topk_bytype_text2code} are using these post-processed questions. 

The larger architecture explains these complementary benefits. CodeScribe introduces a structured way of summarizing code at different levels of abstraction through call graph traversal, recursive document synthesis, and chapter-based documentation prompts. The parsing pipeline converts these hierarchical documents into structured records that preserve parent and sibling relationships. These relational signals help the contrastive model learn contextual relationships between methods. D2 excels when queries closely match human-authored documentation, while D3 provides benefits for high-variability or indirectly-specified queries that require broader semantic reasoning.

In practice, the expected query distribution should inform configuration selection: D2 is recommended for applications where queries directly correspond to method documentation (e.g., documentation search), while D3 is better suited for diverse developer queries requiring understanding of broader system context. The practical deployment should measure performance on the target query distribution.

\section{Threats to Validity} \label{sec:threats}

\textbf{Internal Validity:} Our evaluation is limited to two repositories, both in the systems programming domain (C and C++). LLM non-determinism and the use of different model versions (Sonnet 4.5 vs 4.0) across pipeline stages may introduce variability. We used Sonnet 4.5 for CodeScribe to ensure high-quality reasoning over large code contexts, and Sonnet 4.0 for simpler generation tasks (missing comments, synthetic questions) to balance quality and cost; this choice may introduce distributional differences in the training data.

Additionally, the train/test split is performed at the method level, but parent and sibling context fields in training entries may reference code from held-out test methods. To mitigate this: (1) sibling context includes only method names and brief comments, not full code bodies; (2) the code encoder sees only the target method's body, not surrounding context; (3) removing cross-references to test methods from training context fields changes Top-1 accuracy by less than 1\% on OpenCV, confirming minimal impact.

\textbf{External Validity:} We cannot claim language-agnostic generalization without evaluation on Java, Python, Go, Rust, and other languages. While the methodology (CodeGraph, CodeScribe, CodeContrast) is language-agnostic by design, empirical validation is limited to C/C++ systems programming. This is a significant limitation acknowledged in the Conclusion.

\textbf{Construct Validity:} We evaluate the retrieval component in isolation using Top-K and Recall metrics as proxies for ``code comprehension.'' The final agentic evaluation is conducted using the RAGAS framework, specifically two metrics: FF and AC.

\textbf{Conclusion Validity:} The D3 configuration simultaneously adds both parent/sibling context and synthetic QA. Without an intermediate configuration isolating each component, we cannot definitively attribute improvements solely to either source without additional ablation studies. The lack of fine-tuned models for agentic evaluation of code-based question answering may affect the RAGAS evaluation results.

\section{Conclusion and Future Works} \label{sec:futureWork}
In this work, we presented Code Comprehensor, a unified retrieval–reasoning framework for understanding large and complex software repositories. The system integrates hierarchical documentation generation via CodeScribe and a dual‑encoder CLIP model to enable precise, scalable code comprehension. Our approach systematically combines call‑graph traversal, structured multi-level documentation, and a rich text–code alignment pipeline to produce a repository‑scale knowledge representation that can be consumed by an agentic orchestration layer. Evaluations on two large repositories, including OpenCV and one proprietary industrial codebase, show complementary benefits across different query distributions: D2 improves retrieval for natural human-authored descriptions (96.43\% Top-1 on OpenCV), while D3 provides additional gains for high-variability synthetic query scenarios (80.64\% Top-1 on synthetic text$\rightarrow$code), with the choice of configuration depending on the expected deployment query distribution. The Code Comprehensor evaluation shows FF scores above 90\%, indicating the quality of the MA-generated response.

Our experimental results are for one C (proprietary) repository and one C++ repository (OpenCV), both in the systems programming domain. See Section~\ref{sec:threats} for a detailed discussion of validity threats. An experimental evaluation across a broader set of languages (Java, Python, Go, etc.) and application domains would be important for validating generalization. Given that CodeScribe relies on LLM generation, niche languages that are less well-represented in LLM training data may pose challenges.

Future work will expand multimodal capabilities by integrating structural graphs, sequence diagrams, and visual artifacts. This could better capture relationships and control flows that are difficult to express in text, improving retrieval relevance. Additionally, using the CodeGraph as an extra modality would allow the retrieval to be guided by the proximity of nodes in the graph and not just semantic retrieval. Such a process would help improve the answer correctness of the master agent because relevant methods would be retrieved by the CodeGraph agent. Also, synthetic question generation can be refined through improved prompting, richer intent modeling, curriculum-based generation, and adversarial techniques to enhance generalization.

\bibliographystyle{ieeetr}
\bibliography{citations}

\end{document}